\documentclass[prb,twocolumn,floatfix,preprintnumbers,amsmath,amssymb,superscriptaddress]{revtex4}

\usepackage{graphicx}
\usepackage{amsmath}
\usepackage{dcolumn}
\usepackage{bm}
\usepackage{color}
\usepackage{xspace}
\usepackage{float}

\begin{document}

\title{Magnetic ordering in the ultra-pure site-diluted spin chain materials SrCu$_{1-x}$Ni$_x$O$_2$}

\author{G. Simutis}
 \email{gsimutis@phys.ethz.ch}
\affiliation{Neutron Scattering and Magnetism, Laboratory for Solid State
Physics, ETH Z\"urich, Z\"urich, Switzerland}

\author{M. Thede}
\affiliation{Neutron Scattering and Magnetism, Laboratory for Solid State
Physics, ETH Z\"urich, Z\"urich, Switzerland}

\author{R. Saint-Martin}
\affiliation{Synth\`ese, Propri\'et\'es et Simulation des Mat\'eriaux, SP2M-ICMMO, UMR-CNRS 8182, Universit\'e Paris-Sud, Universit\'e Paris-Saclay, 91405 Orsay Cedex, France}

\author{A. Mohan}
\affiliation{Leibniz Institute for Solid State and Materials Research IFW Dresden, P.O. Box 270116, D-01171 Dresden, Germany}

\author{C. Baines}
\affiliation{Laboratory for Muon Spin Spectroscopy, Paul Scherrer Institut, CH-5232 Villigen, Switzerland}

\author{Z. Guguchia}
\affiliation{Laboratory for Muon Spin Spectroscopy, Paul Scherrer Institut, CH-5232 Villigen, Switzerland}

\author{R. Khasanov}
\affiliation{Laboratory for Muon Spin Spectroscopy, Paul Scherrer Institut, CH-5232 Villigen, Switzerland}

\author{C. Hess}
\affiliation{Leibniz Institute for Solid State and Materials Research IFW Dresden, P.O. Box 270116, D-01171 Dresden, Germany}

\author{A. Revcolevschi}
\affiliation{Synth\`ese, Propri\'et\'es et Simulation des Mat\'eriaux, SP2M-ICMMO, UMR-CNRS 8182, Universit\'e Paris-Sud, Universit\'e Paris-Saclay, 91405 Orsay Cedex, France}

\author{B. B\"uchner }
\affiliation{Leibniz Institute for Solid State and Materials Research IFW Dresden, P.O. Box 270116, D-01171 Dresden, Germany}

\author{A. Zheludev}
\affiliation{Neutron Scattering and Magnetism, Laboratory for Solid State
Physics, ETH Z\"urich, Z\"urich, Switzerland}

\date{\today}

\begin{abstract}
Muon spin rotation technique is used to study magnetic ordering in ultra-pure samples of SrCu$_{1-x}$Ni$_x$O$_2$, an archetypical $S=1/2$ antiferromagnetic Heisenberg chain system with a small amount of $S=1$ defects. The ordered state in the parent compound is shown to be highly homogeneous, contrary to previous report [M. Matsuda et al., Phys. Rev. B \textbf{55}, R11953 (1997)]. Even minute amount of Ni impurities result in inhomogeneous order and a decrease of the transition temperature. At as little as $0.5$~\% Ni concentration, magnetic ordering is entirely suppressed. The results are compared to previous theoretical studies of weakly coupled spin chains with site-defects.
\end{abstract}

 \maketitle
\section{Introduction}

Impurities and defects often have a profound effect on the ground states and excitations in low dimensional quantum magnets.\cite{Dasgupta1980,Fisher1994,Damle2000,Oseroff1995} Due to topology, this is particularly true in one dimension. In gapped spin chains and ladders, where spin correlations are intrinsically short-ranged, defects may in some cases {\it enhance} magnetism by releasing new $S=1/2$ degrees of freedom.\cite{Affleck1987,Affleck1988} The latter may behave as paramagnetic spins,\cite{Hagiwara1990} bind into dimers across their connecting chain fragments,\cite{Zorko2002}  or form complex interacting one-dimensional systems of their own.\cite{Schmidiger2016} In the presence of weak 3-dimensional interactions, the emergent free spins may even undergo long-range magnetic ordering.\cite{Hase1993,Uchiyama1999} In contrast, for gapless spin chains such as the $S=1/2$ Heisenberg model, defects of a certain type tend to disrupt the intrinsic quasi long range correlations, suppressing magnetism.\cite{Eggert2002,Eggert2003,Eggert2004,Eggert2006} In particular, they lead to a drastic reduction of density of states at low energies\cite{Simutis2013} and have a negative impact on magnetic heat transport.\cite{Hlubek2011,Mohan2014} For weakly-coupled gapless spin chains, defects are expected to {\it suppress} long range magnetic order, rather than enhance it.\cite{Eggert2004} The ordering temperature $T_N$, for instance, is predicted to be reduced five-fold, with the introduction of a small number of $S=0$ defects into weakly coupled Heisenberg $S=1/2$ chains.\cite{Eggert2002} It is the theoretically predicted effect of defects on $T_N$ that the present work aims to test experimentally.

Ideal candidates for such a study are the large-$J$  Heisenberg AF $S=1/2$ chain materials Sr$_2$CuO$_3$ and SrCuO$_2$. These compounds exhibit excellent one-dimensional behavior across a broad range of temperatures.\cite{Motoyama1996} While they do order antiferromagnetically in three dimensions at low temperatures due to residual inter-chain interactions, the ordered moment remains very small.\cite{Kojima1997,Matsuda1997,Zaliznyak1999} The most suitable technique to probe such weakly ordered states is muon spin rotation ($\mu$SR).\cite{Yaouanc2011} Unfortunately, introducing and controlling spin-defects in these compounds is far from straightforward.\cite{Hlubek2012,Kojima2004} One problem is that Zn$^{2+}$, a traditional $S=0$ substitute for Cu$^{2+}$, does not readily enter the crystal structure during synthesis.\cite{Kojima2004} To date, only one focused study of long-range ordering in site-diluted Sr$_2$CuO$_3$ has been reported. It was  based on Pd$^{2+}$ substitution, but the exact concentration of defects was shown to be difficult to control, and no systematic studies for the concentration dependence of $T_N$ was performed.

\begin{figure}
\centering
\includegraphics[width={\columnwidth}]{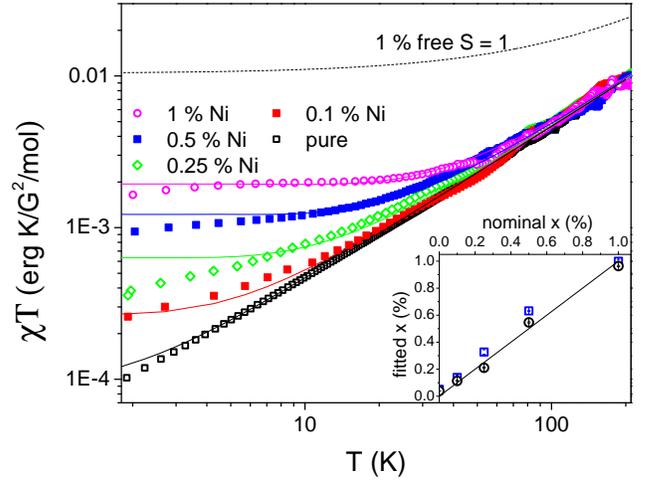}
\caption{Magnetic susceptibility multiplied by temperature for all the studied compounds displayed in a log-log plot for a comparison with the theory of Sirker et al.\cite{Sirker2008} The measured susceptibility follows the prediction very well down to about 10 K, whereupon there are some deviations as discussed in the text.  The dashed line shows the expected response of 1 \% free S = 1 spins in an intact chain. The extracted impurity concentration is shown in the inset for fit using all the data (black circles) and measurements only above 10K (blue squares). Solid black line shows the ideal case of obtained number of impurities equal to the nominal concentration. \label{fig:susceplog}}
\end{figure}

\begin{figure}
\centering
\includegraphics[width={\columnwidth}]{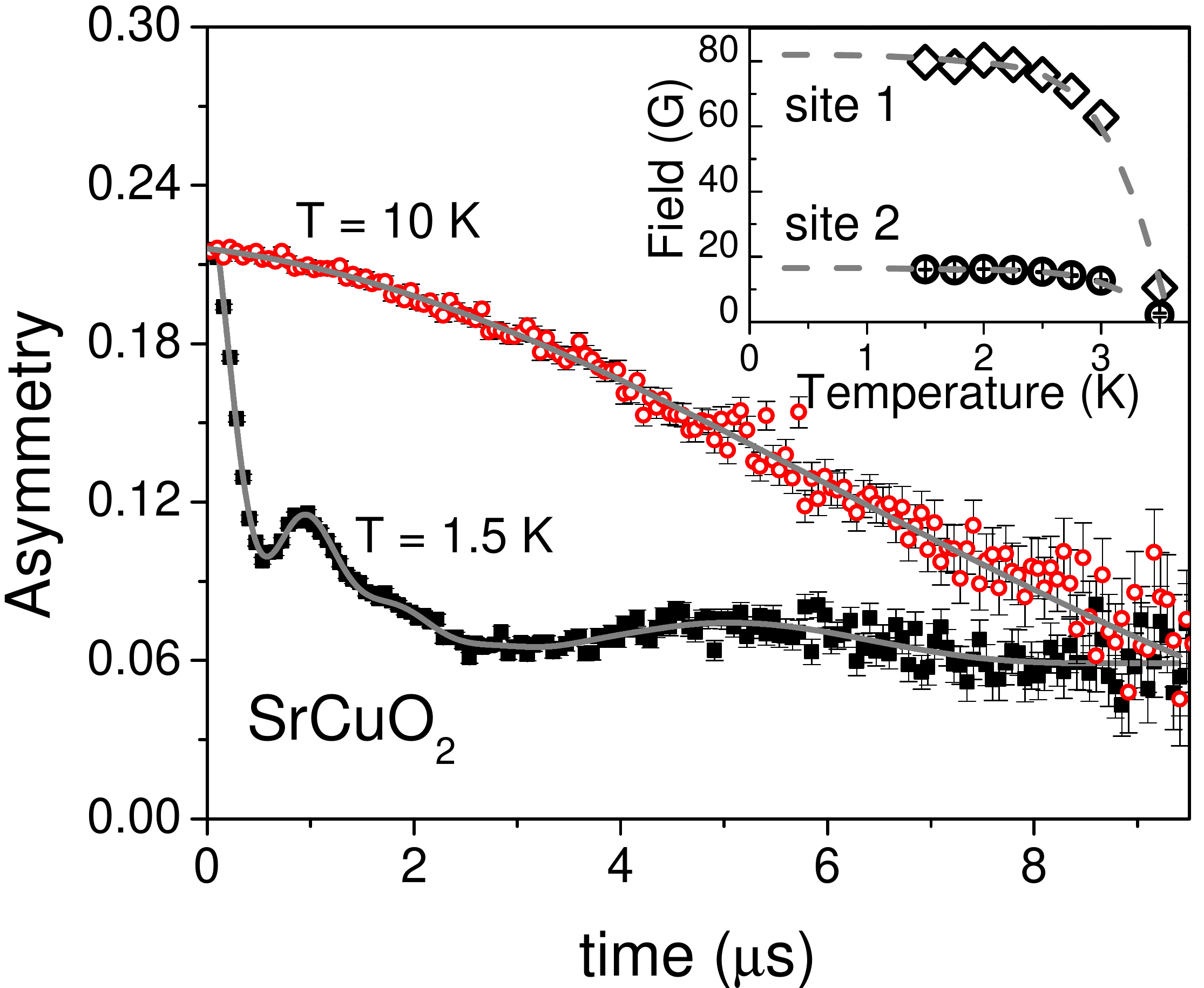}
\caption{Muon decay asymmetry for the pure SrCuO$_2$ above and below the ordering temperature. The solid grey lines are fits to the data as described in the text. The inset shows the temperature dependence of the average internal field in the two muon stopping sites. The dashed lines are guides to the eye. \label{fig:TZFpure}}
\end{figure}

In the present work we focus on defects in  SrCuO$_2.$\cite{Motoyama1996,Matsuda1997} Instead of using Zn$^{2+}$ or Pd$^{2+}$, we introduce defects by Ni$^{2+}$ substitution on the Cu$^{2+}$ sites. Although Ni$^{2+}$ is an $S=1$ ion, its spin becomes screened by quantum fluctuations in the $S=1/2$ host chains. As a result, each $S=1$ impurity acts as a chain break.\cite{Eggert1992, Simutis2013, Karmakar2015} The advantage is that the techniques for precisely controlling Ni$^{2+}$ substitution and ensuring exceptional sample purity in SrCuO$_2$ have been very well established in the course of recent transport studies.\cite{Hlubek2010,Hlubek2011,Hlubek2012,Mohan2014,SaintMartin2015} Below we report on a series of $\mu$SR experiments on this type of ultra-pure samples. From these experiments, we draw several important conclusions regarding long-range order in the parent compound and in the Ni-diluted derivatives: i) contrary to previous reports of glassy behavior,\cite{Matsuda1997} magnetic order in stoichiometric SrCuO$_2$ is a conventional, entirely homogeneous state; ii) A Ni concentration of as little as 0.5\% almost entirely suppresses static magnetic order and iii) the concentration dependence of $T_N$ for low impurity levels is in good agreement with the theoretical predictions of Ref.~\onlinecite{Eggert2002}.

\section{Experiment}

\subsection{Crystal growth and characterization}

Single crystals of SrCu$_{1-x}$Ni$_x$O$_2$ were grown using the floating zone method.\cite{SaintMartin2015}  For pure and Ni-doped SrCuO$_2$ single crystal growth, feed rods with 0.0,  0.1, 0.25, 0.5 and 1 $\%$ of Ni doping concentrations were prepared using SrCO$_3$ (99.99$\%$ or 99.9995$\%$, Alfa Aesar), CuO (99.99$\%$ or 99.9995$\%$, Chempur)  and NiO (99.99$\%$, Alfa Aesar) powders. The powders were ground and sintered several times at 900-990$^{\circ}$C and shaped in a rubber tube (diameter = 6 mm, length = 80 mm) at a pressure of 2500 bar. Subsequently, the cylindrical feed rods were sintered at 995$^{\circ}$C for 24 h in order to obtain very dense rods which are necessary to initiate and maintain the growth experiment. The growth was performed under an O$_2$ atmosphere with a growth rate of 1 mm/h.

The structures of Ni-doped SrCuO$_2$ single crystals were verified by x-ray diffraction (XRD) using a Panalytical X'Pert MPD Pro powder diffractometer and Cu K-alpha 1 radiation. The XRD measurements of powder samples obtained by grinding a part of the single crystals show that the crystals have orthorhombic structures (Cmcm). Their lattice constants, obtained by the Rietveld analysis, does not change significantly in our doping region.\cite{SaintMartin2015}

The magnetic susceptibility measurements were performed on the same crystals that were used in the following muon experiments. The crystals were aligned with the \textbf{c} crystal axis parallel to the applied magnetic field. Vibrating Sample Magnetometer option of the Quantum Design PPMS system was used to take the measurements.

\subsection{Muon spin rotation technique}

This study was performed on the LTF, GPS and GPD muon spectrometers at the Paul Scherrer institute. In order to probe the magnetic fields in all muon sites, the crystals were crushed and then pressurized into pellets. They were attached either to a silver plate (LTF instrument), a fly-through sample holder (for GPS instrument) or placed in a CuBe pressure cell (for GPD instrument). Most of the measurements were performed in zero-field configuration. Additional longitudinal field measurements up to a field of 120 G were performed on the 0.1\% Ni-doped sample. Spectra with applied transverse fields of 15 and 30 G were obtained for the pure as well as 0.1, 0.25 and 0.5\%  Ni-doped samples. Temperatures down to 20 mK were used to measure the samples. Data treatment was performed using MUSRfit program.\cite{Suter2012}

\section{Experimental Results}

\subsection{Magnetic susceptibility of diluted chains}

Since the aim of the present paper is to study the efects of chain fragmentation, it is important to make sure that the chemical substitution has indeed the desired effect of breaking the chains. Figure~\ref{fig:susceplog} shows the susceptibility for all the measured compounds together with a comparison with a response from a corresponding amount of unscreened S=1 ions in an intact chain. It is immediately clear that the measured susceptibility is significantly lower than what would be expected if Ni$^{2+}$ simply behaved as a free spin. Instead, the implanted S = 1 is screened and the only magnetically effective elements are the spin chains of finite length. The dominant contribution to the susceptibility is due to the free S = 1/2 spins\cite{Eggert1992} at the ends of the odd-length chains, however a full description of the magnetic response also includes the boundary susceptibility\cite{Fujimoto2004} and the small contribution coming from the actual fragments of the chains. Such problem has been studied in detail by Sirker et al. in ~\onlinecite{Sirker2008} and the complete expression of the susceptibility per mole reads:

\begin{eqnarray}
\chi (x,T) =\frac{N_A (g\mu_B)^2}{k_B} \big[ \frac{x}{4T} \frac{1-x}{2-x} [1- (1-x)^\frac{J}{T} ] \\ \nonumber
+(1-x)^\frac{J}{T}[(1-x + \frac{xJ}{T})\chi_\mathrm{bulk} + x\chi_\mathrm{bound}] \big]
\end{eqnarray}
In the above equation, $\chi_\mathrm{bound}$ corresponds to the boundary susceptibility and can be expressed\cite{Fujimoto2004} as $1/\chi_\mathrm{bound} = 12Tln(2.9 J/T)]$. In order to properly describe the contribution from the bulk spin chains $\chi_\mathrm{bulk}$, an effective field theory expression\cite{Eggert1994} was used for T $<$ 0.01 J and a QMC-derived function\cite{Johnston2000} for higher temperatures. A region of 5 K around this value was described as a linear interpolation between the two expressions.

The obtained data is described by this picture very well as can be seen in Fig.~\ref{fig:susceplog}. A discrepancy at very low temperatures is observed, where the measured susceptibility is below the expected one. Such discrepancy has also been noted in measurements done previously\cite{Karmakar2015} and interpreted as arising due to three-dimensional interactions. The number of chain breaks $x$ can be fitted to the expression above and compared to the nominal composition as shown in the inset of Fig:~\ref{fig:susceplog}. While small deviations persist and there is a minor dependency of the extracted parameters on the fitting range, the susceptibility data is in general very well described by the fragmented chain model. Moreover, it is evident that the desired amount of spin breaks is obtained by Ni-substitution.

\subsection{Magnetic order in pure SrCuO$_2$}

\begin{figure}
\centering
\includegraphics[width={\columnwidth}]{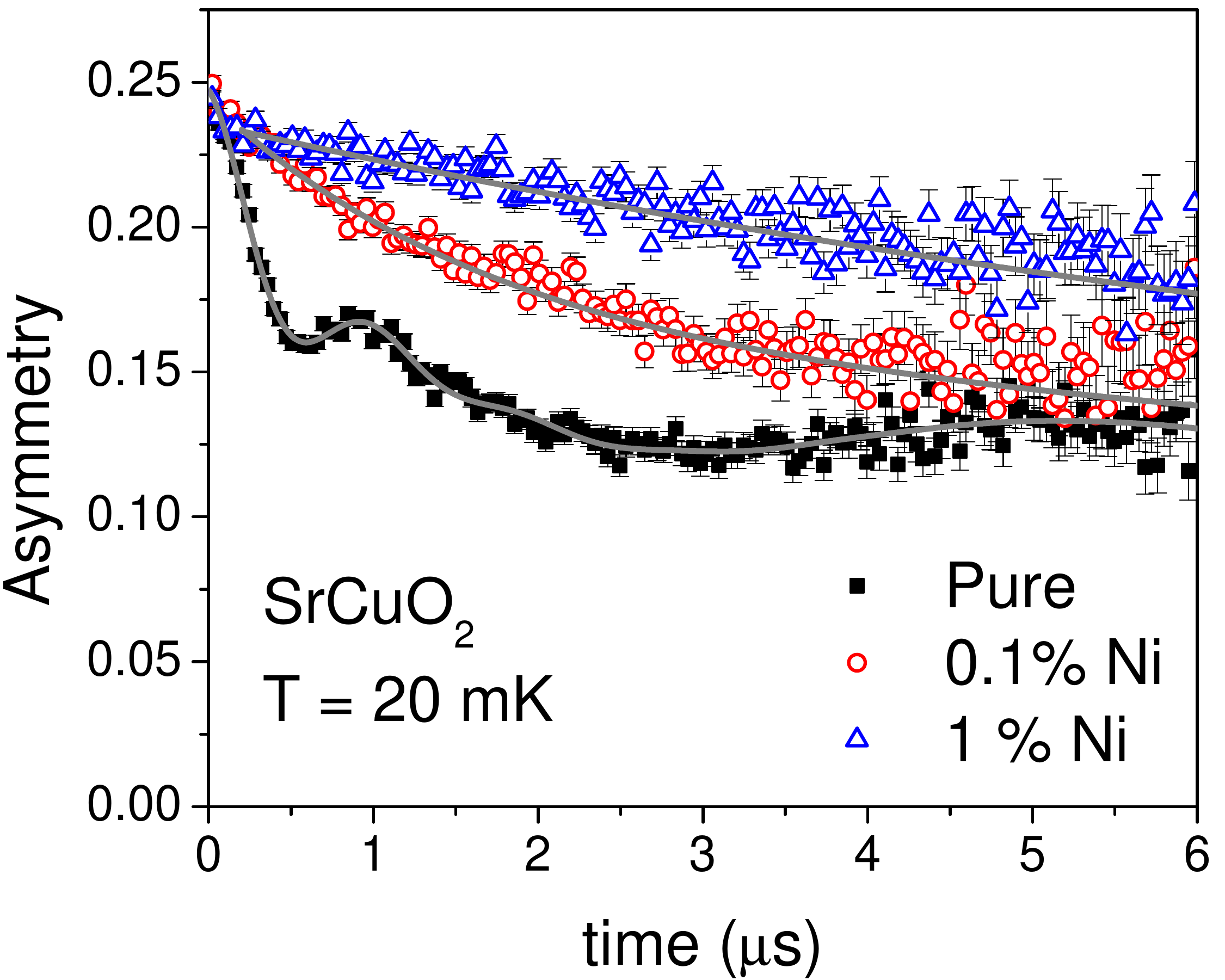}
\caption{Muon spin rotation asymmetry for the clean and Ni-substituted SrCuO$_2$. The oscillations are suppressed already in the sample with only 0.1$\%$ Ni impurities, which suggests inhomogeneous ordering. The increased background compared to Fig.~\ref{fig:TZFpure} is due to the use of  a different sample environment (dilution refrigerator). \label{fig:TZFdoped}}
\end{figure}

As a first step, $\mu$SR measurements were done on the stoichiometric parent compound. As noted, earlier studies have not detected clear signs of homogeneous magnetic order, but only an increase of relaxation at low temperatures\cite{Matsuda1997}. In contrast, the low-$T$ spectra collected on our ultra-pure samples (Fig.~\ref{fig:TZFpure}) show well defined oscillations. This is a clear indication of homogeneous static magnetic order.  As expected, above the phase transition, the $\mu$SR asymmetry is well described by a Kubo Toyabe type relaxation which  is typical of paramagnetic materials and originates from the interaction of the muon spin with randomly oriented nuclear magnetic moments.\cite{Yaouanc2011} To describe the oscillations observed at low temperatures, using several $\cos$-functions yields unphysical 45$^\circ$ phase shifts. After considering several models, we found that the data are best reproduced by two Bessel functions, corresponding to two distinct muon stopping positions in the unit cell:
\begin{eqnarray}
A(t) &=& \frac{2}{3}\left[ A_1 J_0 (\gamma_\mu B_1 t)\exp(-\lambda_1 t \right) \nonumber \\
&+&  A_2 J_0 (\gamma_\mu B_2 t) \exp(-\lambda_2 t) ] \nonumber \\
&+& \frac{1}{3} (A_1 + A_2) \exp(-\lambda_{tail}t)
\end{eqnarray}
The resulting fits for the low temperature Muon decay asymmetry of the pure material are shown in solid lines in Fig.~\ref{fig:TZFpure} and Fig.~\ref{fig:TZFdoped}. Bessel-type oscillations are indicative of incommensurate magnetic structures,\cite{Yaouanc2011} which is fully consistent with previous neutron scattering studies.\cite{Zaliznyak1999} The measured temperature dependence of the average internal magnetic field at the two muon sites is shown in the inset of Fig.~\ref{fig:TZFpure}, and can be viewed as a plot of the magnetic order parameter. Two distinct precession frequencies occur in the $\mu$SR spectra, corresponding to the local magnetic fields of 80 G (75\% of the signal) and 16 G (25\% of the signal). Note that the observed ordering temperature in our samples is higher than the 2 K value reported in Ref.~\onlinecite{Matsuda1997}.

\subsection{Magnetic order in SrCuO$_2$ with Ni impurities}

\begin{figure}
\centering
\includegraphics[width={\columnwidth}]{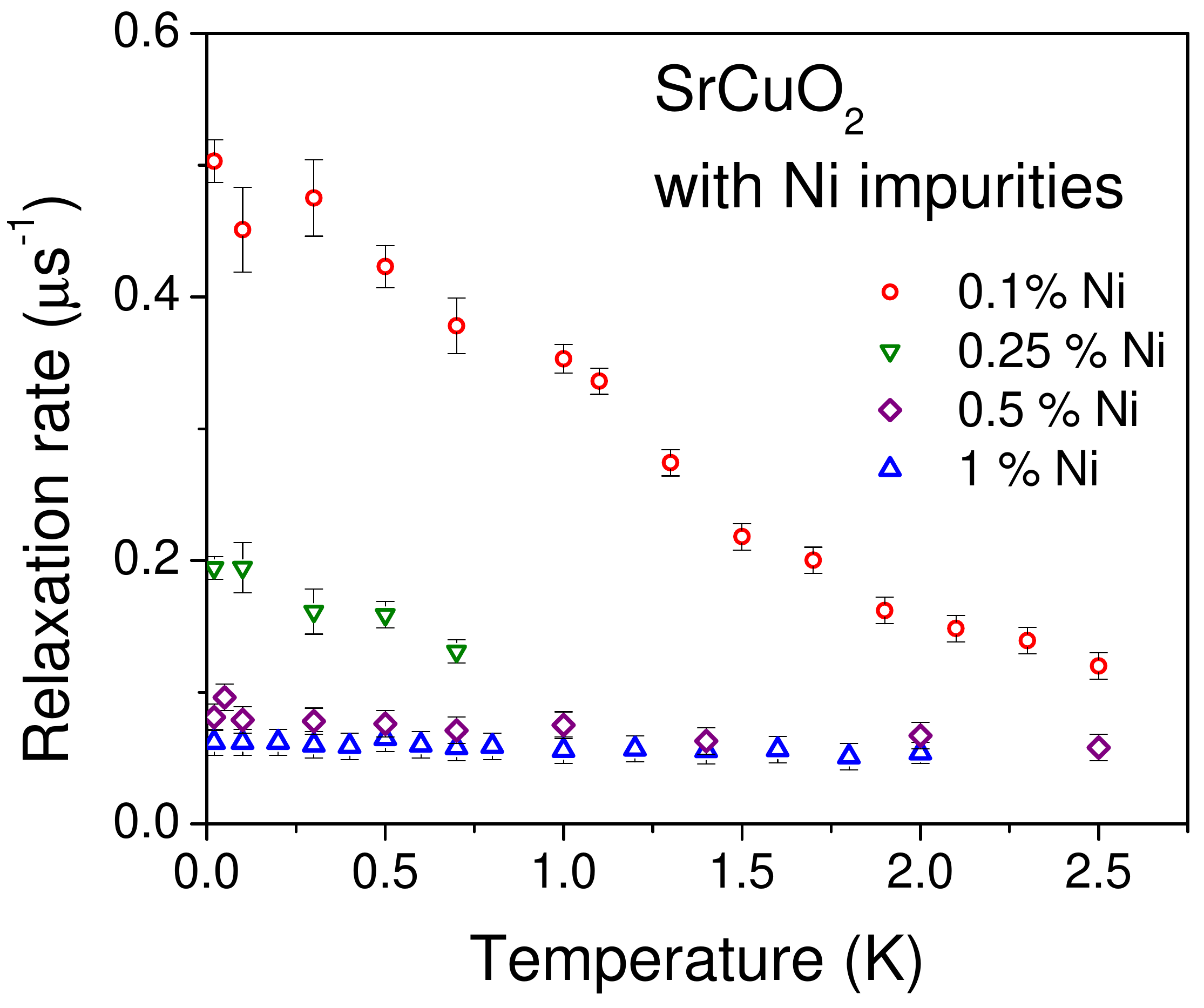}
\caption{Extracted relaxation rate in the doped versions of SrCuO$_2$. The obtained data points are from an empirical fit of a zero field spectrum to a simple exponential relaxation.\label{fig:Lambda}}
\end{figure}

\begin{figure}
\centering
\includegraphics[width={\columnwidth}]{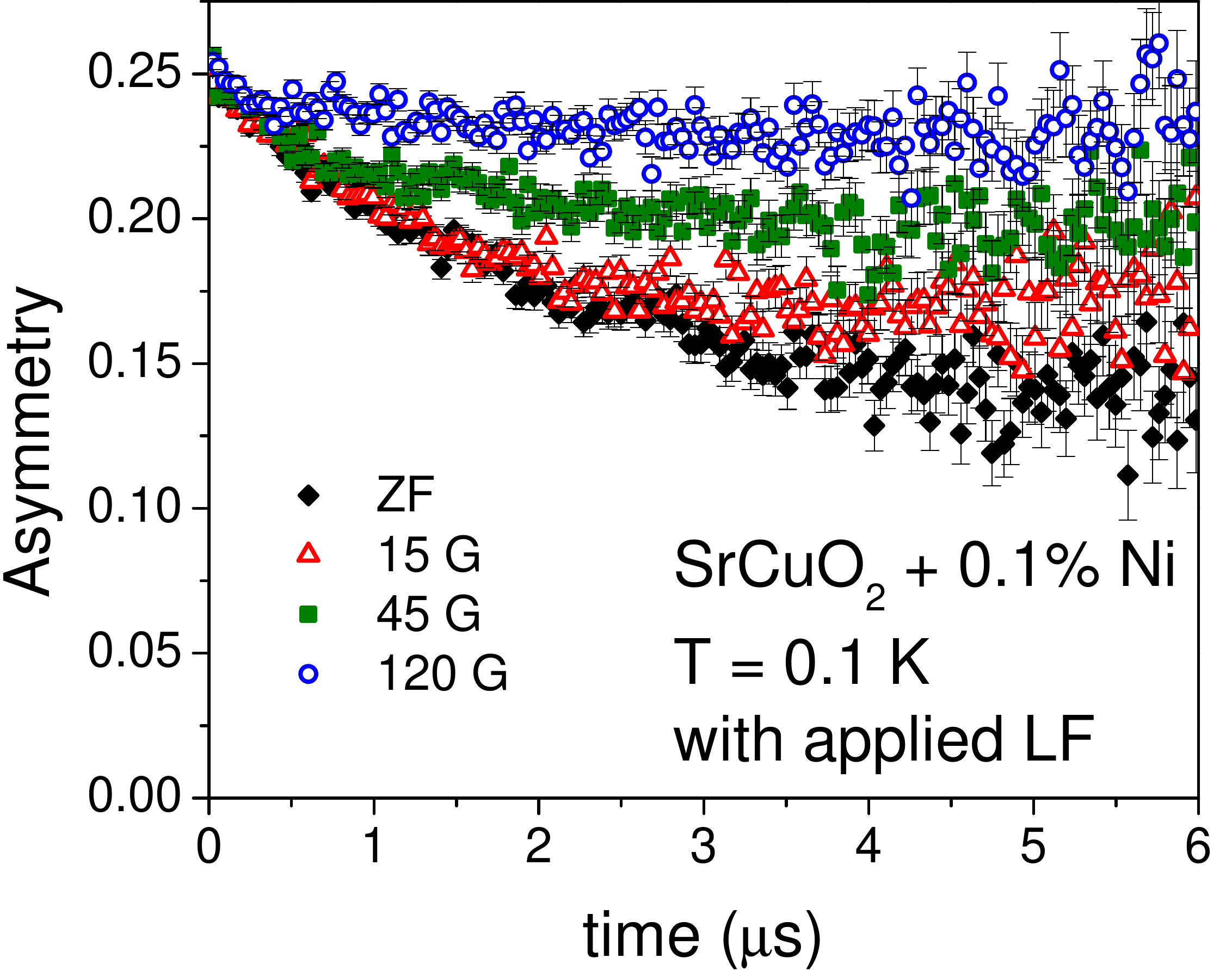}
\caption{Muon decay asymmetry spectra for SrCuO$_2$ with 0.1$\%$ Ni impurities under an applied longitudinal magnetic field. The asymmetry recovers almost fully with an applied field of 120 G indicating static magnetic order in the sample.
\label{fig:LFdoped}}
\end{figure}

Oscillatory relaxation behavior has not been observed in any of our Ni-substituted samples (Fig.~\ref{fig:TZFdoped}). Instead, at all temperatures, the relaxation is described by a simple exponential form (Fig.~\ref{fig:TZFdoped}, for 0.1\% and 1\%-Ni samples). For samples with Ni-content $x\le 0.25$~\% the measured relaxation rate increases substantially upon cooling to low temperatures (Fig.~\ref{fig:Lambda}). This behavior is indicative of either i) a progressive slowing down of dynamic spin fluctuations or ii) the onset of static but highly inhomogeneous magnetic order. The two cases can be distinguished in longitudinal-field experiments. In the 0.1\%-Ni sample, the decay asymmetry was found to almost fully recover under a weak longitudinal field (Fig.~\ref{fig:LFdoped}), confirming that the magnetic order is indeed static. For Ni-content $x\gtrsim 0.5$~\% the relaxation rate measured in zero field experiments is totally temperature-independent. In this case, magnetic ordering appears to be totally suppressed.

\begin{figure}
\centering
\includegraphics[width={\columnwidth}]{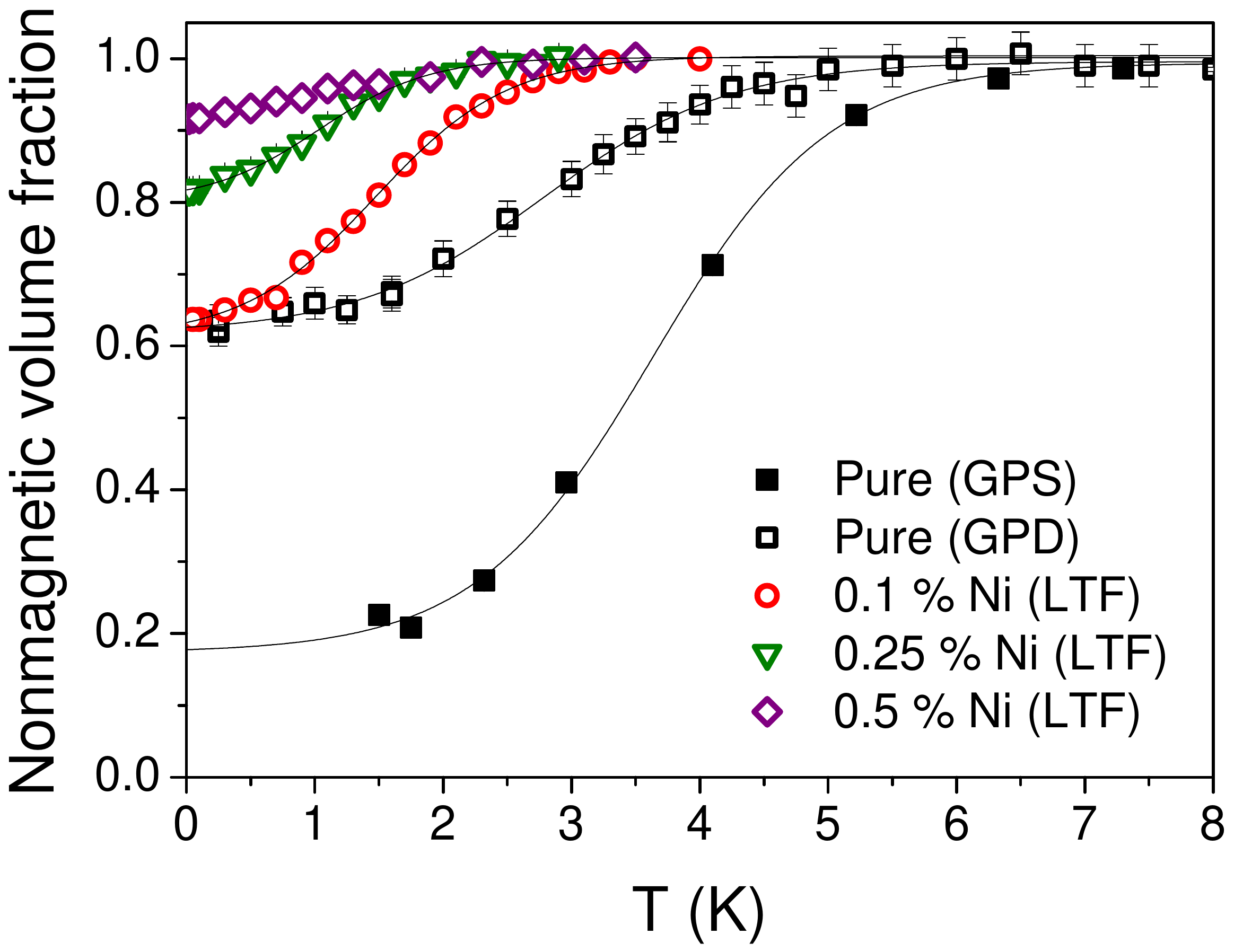}
\caption{Nonmagnetic volume fraction obtained from transverse field measurements of pure and the site-diluted SrCuO$_2$. The pure SrCuO$_2$ was measured using the low background GPS instrument as well as the higher-background GPD instrument. The compounds with impurities were measured using the LTF spectrometer which also has a considerable background. The lines are fits to the sigmoid function as described in the text.\label{fig:WTFraw}}
\end{figure}

\begin{figure}
\centering
\includegraphics[width={\columnwidth}]{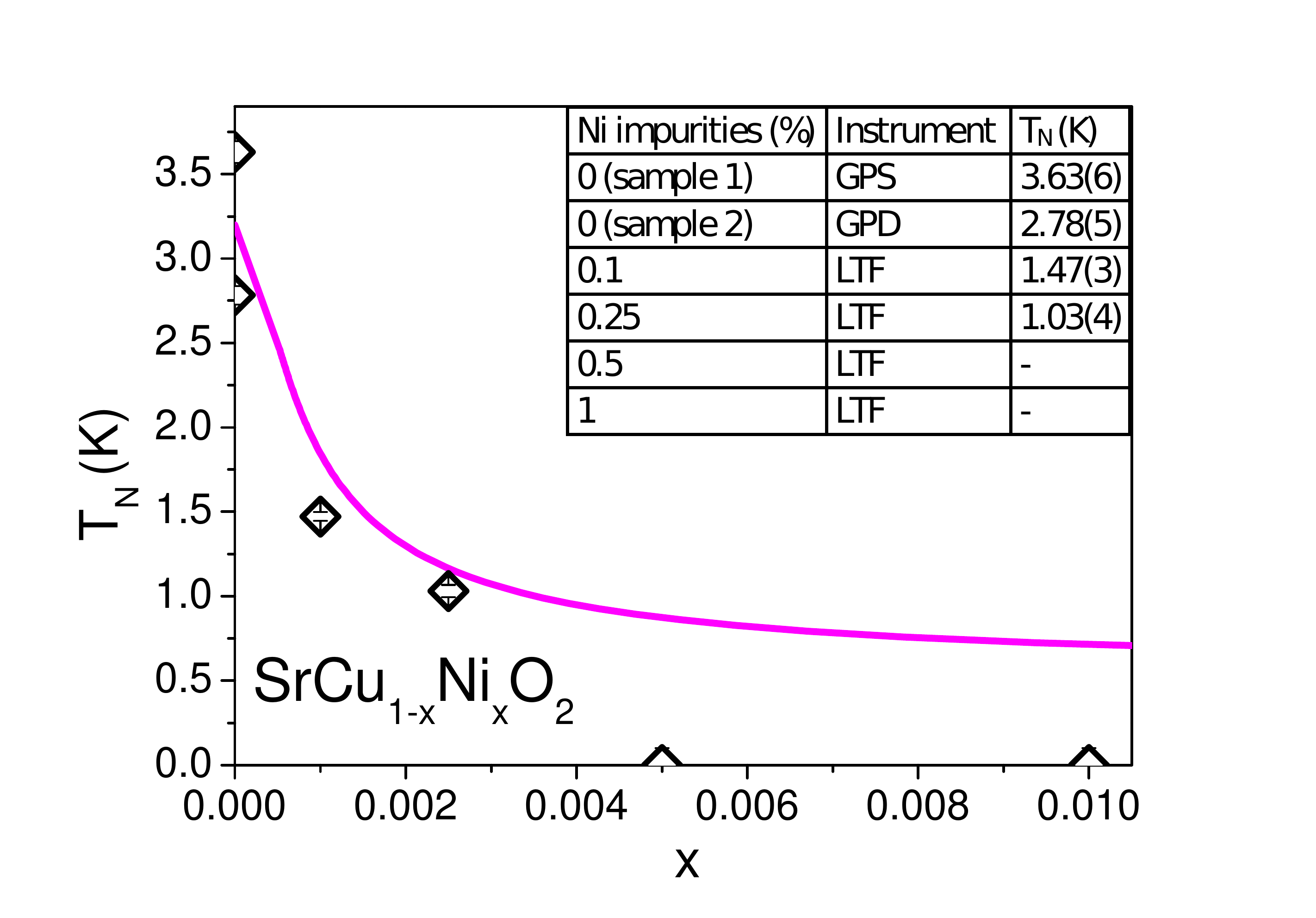}
\caption{Ordering temperatures extracted from the transverse field measurements. The solid line is the calculation from Ref.~\onlinecite{Eggert2002}. While the ordering temperature in the samples with low concentration of impurities follows the predicted behaviour, ordering was found to be completely suppressed in the samples with 0.5\% and 1\% impurities.\label{fig:TN}}
\end{figure}

A more accurate determination of the ordering temperature was obtained in transverse-field muon experiments that are a tool to probe the magnetic volume fraction.\cite{Yaouanc2011} The non-magnetic volume fraction measured in several of our samples in an applied transverse field of 30~G is plotted against temperature in Fig.~\ref{fig:WTFraw}. Note that the plotted volume fraction is not normalized in the sense that it includes contributions from both the samples and the sample environment. As a result, it never goes to zero at low temperature, even in a perfectly ordered system.  Additionally, even in the very low background GPS instrument, the {\it apparent} non-magnetic volume fraction remains finite due to one of the muon sites experiencing very weak internal field (as seen in the inset of Fig.~\ref{fig:TZFpure}). To extract the transition temperature, the volume fraction measurements for $x\lesssim 0.25$~\% were fit to empirical sigmoidal functions (Fig.~\ref{fig:WTFraw}, symbols). The central value was associated with $T_N$ and is plotted vs. Ni-concentration in Fig.~\ref{fig:TN}. A ``zero'' value of $T_N$ for $x=0.5$~\% and $x=1$~\% indicates a lack of magnetic muon spin depolarization in zero field. Note that the data include two distinct pure SrCuO$_2$ samples which appear to have slightly different ordering temperatures.

\section{Discussion}

Previous studies of magnetic ordering in SrCuO$_2$ have left many unanswered questions, notably why the oscillations in the muon spectrum were absent,\cite{Matsuda1997} as well as why there seemed to be more than one characteristic temperature observed in the neutron study.\cite{Zaliznyak1999} Our present investigation suggests that both issues can be attributed to sample quality. We unambiguously show that in ultra high purity SrCuO$_2$ samples the muon spins precess coherently around the static local magnetic fields. Moreover, experiments on the Ni-substituted samples illustrate that even minute concentrations of impurities may disrupt order.  Crystals of even 99.9\% purity are simply not good enough to characterize magnetic transition in the parent compound. In fact, the muon spectra reported in Ref.~\cite{Matsuda1997} for ``$x=0$'', look very similar to the our data for $x=0.001$. Similarly, the large crystals used in the neutron diffraction experiments of Ref.~\cite{Zaliznyak1999} may have have been affected by very small but significant inhomogeneities, resulting in a distribution of ordering temperatures. Even in our ultra high purity SrCuO$_2$ samples we see slightly different transition temperatures in different batches.

The main result of the present study is the measured defect-concentration dependence of $T_N$ (Fig.~\ref{fig:TN}). It is generally in very good agreement with the theoretical predictions of Ref.~\cite{Eggert2002}, represented in Fig.~\ref{fig:TN}  by the solid curve. The same theoretical work specifically predicts a highly inhomogeneous ordered state, also fully consistent with our observations. A similar inhomogeneity manifested in a damping of muon oscillations has also been reported for Sr$_2$CuO$_3$.\cite{Kojima2004}

The clear discrepancy between the predictions of  Ref.~\cite{Eggert2002} and our data is the apparent total absence of magnetic ordering in samples with 0.5 \% and 1\% Ni impurities. Based on the theory, we would have expected magnetic order to survive, and $T_N$ to decrease to approximately $0.6$~K, which is still readily accessible in experiments. Of course, our failure to detect magnetic order may simply be due to a strongly suppressed ordered moment, outside the sensitivity range of our technique. Alternatively, the total suppression of magnetic order may be due to the double-chain structure of SrCuO$_2$ and a natural frustration of inter-chain interactions.\cite{Motoyama1996} Such frustration may also be a reason why the moderately diluted compounds order at slightly lower temperatures than expected.

In the course of preparing the present manuscript we became aware of Ref.~\cite{Karmakar2016} that reports an {\it enhancement} of magnetic ordering upon Co$^{2+}$ substitution in SrCuO$_2$. Note that the case of Co-substitution is a fundamentally different problem from the one considered here. Co$^{2+}$ is a Kramers ion, and therefore does {\it not} lead to an effective breaking of the spin chains the way integer-spin defects such as Ni$^{2+}$ do.\cite{Eggert1992}

\section{Conclusion}

In summary, our experiments demonstrate that 3-dimensional long range ordering in quasi-1-dimensional spin systems is extremely fragile. Minuscule concentrations of magnetic site defects disrupt the homogeneity of the ordered state and even suppress ordering altogether. Once again this emphasizes the importance of using samples of adequate purity in any studies of this kind.

\section{Acknowledgements}
This work was supported by the Swiss National Science Foundation, Division~2. The work at IFW has been supported by the Deutsche Forschungsgemeinschaft through the D-A-CH Project No. HE 3439/12 and the European Commission through the LOTHERM project (Project No. PITN-GA-2009-238475).  GS would like to thank Dr. J. S. M\"oller and Dr. S. Gvasaliya for helpful discussions.

\bibliography{SCOmuSRbib}

\end{document}